\documentclass[aps,prd]{revtex4}%
\usepackage{amsfonts}
\usepackage{amsmath}
\usepackage{amssymb}
\usepackage{graphicx}%
\begin{document}
\title{B=2 Oblate Skyrmions}
\author{L. Marleau and J.-F. Rivard}
\affiliation{D\'{e}partement de physique, g\'{e}nie physique et d'optique, Universit\'{e}
Laval, Qu\'{e}bec QC Canada, G1K 7P4}
\keywords{Skyrme Model, Skyrmions,}
\pacs{PACS numbers: 11.25.Mj, 13.85.Qk }
\email{lmarleau@phy.ulaval.ca}

\begin{abstract}
The numerical solution for the $B=2$ static soliton of the $SU(2)$ Skyrme
model shows a profile function dependence which is not exactly radial. We
propose to quantify this with the introduction of an axially symmetric oblate
ansatz parametrized by a scale factor $d.$ We then obtain a relatively
deformed bound soliton configuration with $M_{B=2}/M_{B=1}=1.958$. This is the
first step towards to description of $B>1$ quantized states such as the
deuteron with a non-rigid oblate ansatz where deformations due to centrifugal
effects are expected to be more important.

\end{abstract}
\volumeyear{year}
\volumenumber{number}
\issuenumber{number}
\eid{identifier}
\date{\today}
\startpage{1}
\endpage{ }
\maketitle
\tableofcontents

\section{}

\section{INTRODUCTION}

The Skyrme model \cite{Skyrme} is a nonlinear effective field theory of weakly
coupled pions in which baryons emerge as localized finite energy soliton
solutions. The stability of such solitons is guaranteed by the existence of a
conserved topological charge interpreted as the quantum baryon number $B$.
More specifically, Skyrmions consist in static pion field configurations which
minimize the energy functional of the Skyrme model in a given nontrivial
topological sector. The model is partly motivated by the large-$N_{c}$ QCD
analysis \cite{t'Hooft,Witten}, as there are reasons to believe that once
properly quantized, a refined version of the model could accurately depict
nucleons as well as heavier atomic nuclei with mesonic degrees of freedom
\cite{Braaten, Houghton, Scoccola} in the low energy limit.

In the lowest nontrivial topological sector $B=1$, the Skyrmion is described
by the spherically symmetric hedgehog ansatz which reproduces experimental
data with an accuracy of $30\%$ or better \cite{Adkins}. However, this
relative success radically contrasts with the situation encountered in the
$B>2$ sector, where the hedgehog ansatz is not the lowest energy configuration
and would not give rise to bound state configurations \cite{Jackson,Weigel}.
Moreover, pioneering numerical investigations of Verbaarschot
\cite{Verbaarschot} clearly indicate that the $B=2$ Skyrmion is not
spherically symmetric, but rather possesses an axial symmetry reflected in its
doughnut-like baryon density. Further inspection of this numerical solution,
in particular the profile function, suggests that the classical biskyrmion may
be represented by an oblate field configuration. Yet, most of the trial
functions used to describe such a solution assume a decoupling from the
angular degrees of freedom, i.e. $F(r,\theta,\phi)=F(r)$, as this is the case
for the instanton-inspired ansatz proposed in \cite{Atiyah} or in the early
variational approach \cite{Thomas,Kurihara} for example. On the other hand,
some axially symmetric solutions were analysed in the $B=1$ sector, by
\cite{Hajduk} and \cite{Marleau} respectively, to include possible
deformations due to centrifugal effects undergone by the rotating Skyrmion and
account for the quadrupole deformations of baryons. Here, our aim is to extend
the work on oblate Skyrmions in \cite{Marleau} to describe dibaryon states.
The classical static oblate solution introduced in this manner will provide a
quantitative estimate of the axial deformation, which is different from a
uniform scaling in a given direction as performed in \cite{Hajduk}. It should
also provide an adequate ground to perform the quantization of the $B=2$ soliton.

There has been several attempts to decribed the angular dependence of the
$B>1$ solutions which is much more complicated than the hedgehog form in
$B=1.$ Fortunately, a few years ago, Houghton et al. \cite{Houghton} came up
with an interesting ansatz based on rational maps. The rational map ansatz
provides a simple alternative compared to the full numerical study of the
angular dependence of the baryon density distribution of multiskyrmions. It
also yields static energy predictions in good agreement with numerical
solutions for several values of $B$. The most interesting feature of this
method remains without doubt that fundamental symmetries of multiskyrmions can
easily be implemented in the ansatz solutions. This provides a clever way to
identify the symmetries of the exact solutions, which are not always apparent,
and in some cases, adequate initial solutions for lengthy numerical
calculations. Close as it may be, the rational map ansatz remains an
approximation and in some cases, more accurate angular ansatz have been found.
For example, Houghton and Krusch \cite{Houghton2} slightly improved the mass
approximation of the biskyrmion by relaxing the requirement of holomorphicity
imposed on rational maps. However, the profile function defined in this work
still solely depends on $r$. As may seems evident, some accuracy still may be
gained by introducing a more appropriate parametrization of the soliton shape
function. Recently, Ioannidou et al. \cite{Ioannidou} obtained similar results
by introducing an improved harmonic maps ansatz where the profile function
depends on radial and polar degrees of freedom as well. However, they had to
deal with a complicated second-order partial differential equation.

From these considerations, we propose a $B=2$ oblate solution based on
rational maps, which could be understood as the rational maps solution
proposed in \cite{Houghton, Krusch} with the radial dependence $F(r)$ replaced
by an oblate form $f(\eta)$. Consequently, the soliton undergoes a smoothly
flattening along a given axis of symmetry. The parameter $d$ provides a
measure of the scale at wich the deformation becomes important while the
solution preserves the angular dependence given by the rational maps scheme
for the $B=2$ case. This choice is obviously consistent with the toroidal
baryon density of the $B=2$ Skyrmion. Implementing the oblate ansatz to the
model, we first integrate analytically the angular degrees of freedom. This
explains why other angular ansatz such as those in \cite{Houghton2} and
\cite{Ioannidou} were not chosen; they led to complications. The second step
involve solving the remaining nonlinear ordinary second-order differential
equation resulting from the minimization of the static energy functional with
respect to the profile function $f(\eta)$. Thereby, the parameter $d$ is set
as to minimize the static energy, i.e. the mass of the soliton. Although the
method applies to higher baryon numbers and other Skyrme model extensions, the
analysis is restricted here to $B=2$ Skyrmion for $SU(2)$ Skyrme model.

In the next section, we present the axially symmetric oblate ansatz for the
SU(2) Skyrme model introducing the oblate spheroidal coordinates. In section
III, we briefly describe rational maps and show how they can be used in the
context of static oblate biskyrmions. A discussion of the numerical results
follows in the last section, where we also draw concluding remarks about how
the oblate-like solution could be a good starting point to perform the
quantization of the non-rigid $B=2$ soliton as the deuteron.

\section{STATIC OBLATE MULTISKYRMIONS}

Let us first introduce the oblate spheroidal coordinates $(\eta,\theta,\phi),$
which are related to Cartesian coordinates through
\begin{equation}
(x,y,z)=d(\cosh\eta\sin\theta\cos\phi,\cosh\eta\sin\theta\sin\phi,\sinh
\eta\cos\theta),
\end{equation}
so a surface of constant $\eta$ correspond to a sphere of radius $d\cosh\eta$
flattened in the $z$-direction by a factor of $\tanh\eta$. For small $\eta$,
these surfaces are quite similar to that of pancakes of radius $d$ whereas
when $\eta$ is large, they become spherical shells of radius given by
$(d/2)e^{\eta}$. Note that taking the double limit $d\rightarrow
0,\eta\rightarrow\infty$ such that $r$ always remains finite, one recovers the
usual spherical coordinates. Thus, the choice of the parameter $d$ establishes
the scale at which the oblateness becomes significant. Finally, the element of
volume reads%

\begin{align}
dV = -d^{3} \cosh\eta\left(  \sinh^{2}\eta+ \cos^{2}\theta\right)
d\eta\ d(\cos\theta) \ d\phi.
\end{align}

Neglecting the pion mass term, the chirally invariant Lagrangian of the
$SU(2)$ Skyrme model just reads
\begin{equation}
{\mathcal{L}}=-\frac{F_{\pi}^{2}}{16}\operatorname*{Tr}L_{\mu}L^{\mu}+\frac
{1}{32e^{2}}\operatorname*{Tr}[L_{\mu},L_{\nu}]^{2}%
\end{equation}
where $L_{\mu}=U^{\dagger}\partial_{\mu}U$ with $U\in SU(2)$. Here, $F_{\pi}$
is the pion decay constant and $e$ is sometimes referred to as the Skyrme
parameter. In order to implement an oblate solution, let us now replace the
hedgehog ansatz
\begin{equation}
U=e^{i\mathbf{\tau}\ \cdot\ \mathbf{\widehat{r}}\ F(r)}%
\end{equation}
by the static oblate solution defined as follow
\begin{equation}
U=e^{i\mathbf{\tau}\ \cdot\ \mathbf{\widehat{\eta}}\ f(\eta)}%
\end{equation}
where the ${\tau}_{k}$ stand for the Pauli matrices while ${\widehat{\eta}}$
is the standard unit vector ${\widehat{\eta}}={\vec{\nabla}\eta}/{|\vec
{\nabla}\eta|}$. More explicitly, this unit vector is simply
\begin{equation}
{\widehat{\eta}}=\frac{(\cosh\eta\sin\Theta\cos\Phi,\cosh\eta\sin\Theta
\sin\Phi,\sinh\eta\cos\Theta)}{\sqrt{\cosh^{2}\eta-\cos^{2}\Theta)}},
\end{equation}
As will become apparent in the next section, we consider the case where
$\Theta\equiv\Theta(\theta)$ and $\Phi\equiv\Phi(\phi),$ i.e. $\Theta$ and
$\Phi$ depend only on the polar angle $\theta$ and the azimuthal angle $\phi$
respectively. Furthermore, $f(\eta)$, which determines the global shape of the
soliton, plays the role of the so-called profile function. In that respect,
the oblate ansatz is clearly different from a scale transformation along one
of the axis \cite{Hajduk}. As in its original hedgehog form, the field
configuration $U$ constitutes a map from the physical space ${\mathcal{R}}%
^{3}$ onto the Lie group manifold of $SU(2)$. Finite energy solutions require
that this $SU(2)$ valued field goes to the trivial vacuum for asymptotically
large distances, that is $U(\eta\rightarrow\infty)\rightarrow1_{2\times2}$.

The expression for the static energy density is
\begin{equation}
{\mathcal{E}}={\mathcal{E}}_{2}+{\mathcal{E}}_{4}=-\frac{F_{\pi}^{2}}%
{16}Tr(L_{i}L_{i})+\frac{1}{32e^{2}}Tr\left(  [L_{i},L_{j}]^{2}\right)
\end{equation}
so, after substituting the oblate ansatz, the mass functional can be written
as
\begin{equation}
M[f(\eta),\Theta(\theta),\Phi(\phi)]=\int dV\left(  {\mathcal{M}}%
_{2}+{\mathcal{M}}_{4}\right)
\end{equation}
with
\begin{equation}
{\mathcal{M}}_{2}=\frac{F_{\pi}^{2}}{8}\left(  |\vec{\nabla}\eta
|^{2}\ f^{\prime2}+\sin^{2}f\ {\tilde{K}}\right)  \label{m2}%
\end{equation}
and
\begin{equation}
{\mathcal{M}}_{4}=\frac{1}{4e^{2}}\left(  2|\vec{\nabla}\eta|^{2}\ f^{\prime
2}\ {\tilde{K}}+\sin^{4}f\ \left(  {\tilde{K}}^{2}-{\tilde{K}}_{ab}{\tilde{K}%
}_{ab}\right)  \right)  . \label{m4}%
\end{equation}
Here, the notation is lighten by the use of the ${\tilde{K}}_{ab}$ matrix
defined as:
\begin{equation}
{\tilde{K}}_{ab}=\vec{\nabla}_{i}\widehat{\eta}_{a}\vec{\nabla}_{i}%
\widehat{\eta}_{b}%
\end{equation}%
\begin{equation}
{\tilde{K}}=Tr({\tilde{K}}_{ab})={\tilde{K}}_{aa}.
\end{equation}
Introducing an auxiliary variable for convenience,
\begin{equation}
\Sigma=\cos^{2}\Theta\ \sin^{2}\Theta+\Theta^{\prime2}\ \cosh^{2}\eta
\ \sinh^{2}\eta\label{par1}%
\end{equation}
one easily deduces that
\begin{equation}
{\tilde{K}}=\left(  \frac{|\vec{\nabla}\eta|^{2} \Sigma}{ (\cosh^{2}\eta
-\cos^{2}\Theta)^{2}} + \Phi^{\prime2}\ \frac{(\sin^{2}\Theta/ \sin^{2}%
\theta)}{d^{2} (\cosh^{2}\eta-\cos^{2}\Theta)}\ \right)  \label{K1}%
\end{equation}
and
\begin{equation}
{\tilde{K}}^{2}-{\tilde{K}}_{ab}{\tilde{K}}_{ab}=\frac{2}{d^{2}}\ \Phi
^{\prime2}\ \frac{|\vec{\nabla}\eta|^{2} \Sigma(\sin^{2}\Theta/ \sin^{2}%
\theta)}{(\cosh^{2}\eta-\cos^{2}\Theta)^{3}}. \label{K2}%
\end{equation}
However, before minimizing the mass functional with respect to the chiral
angle $f(\eta)$, in view to get the static configuration of the soliton, one
must specify an angular dependence in $\Theta(\theta)$ and $\Phi(\phi)$. This
is the subjet of the next section, after a brief recall of some basic features
related to the rational maps ansatz.

\section{OBLATE SKYRMIONS AND RATIONAL MAPS}

Formally, a rational map of order $N$ consists in a ${\mathcal{S}}%
^{2}\rightarrow{\mathcal{S}}^{2}$ holomorphic map of the form
\begin{equation}
R_{N}(z)=\frac{p(z)}{q(z)}%
\end{equation}
where $p$ and $q$ appear as polynomials of degree at most $N$. Moreover, these
maps are built in such a way that $p$ or $q$ is precisely of degree $N$. It is
also assumed that $p$ and $q$ do not share any common factor. Any point $z$ on
${\mathcal{S}}^{2}$ is identified via stereographic projection, defined
through $z=\tan(\frac{\theta}{2})\ e^{i\phi}$. Thus, the image of a rational
map $R(z)$ applied on a point $z$ of a Riemann sphere corresponds to the unit
vector
\begin{equation}
\widehat{n}_{R}=\frac{1}{1+|R_{N}|^{2}}\ \left(  2\Re R_{N}(z),2\Im
R_{N}(z),1-|R_{N}|^{2}\right)
\end{equation}
which also belongs to a Riemann sphere. The link between static soliton chiral
fields and rational maps \cite{Houghton} follows from the ansatz
\begin{equation}
U_{R}(r,z)=e^{i\widehat{n}_{R}\cdot\vec{\tau}F(r)}%
\end{equation}
inasmuch as $F(r)$ acts as radial chiral angle function. To be well defined at
the origin and at $r\rightarrow\infty$, the boundary conditions must be
$F(0)=k\pi$ where $k$ is an integer and $F(\infty)=0.$ The baryon number is
given by $B=Nk$ where $N=\max(\deg p,\deg q)$ is the degree of $R_{N}(z).$ We
consider only the case $k=1$ here, so $B=N$.

By analogy with the nonlinear theory of elasticity \cite{Manton}, Manton has
showed that the static energy of Skyrmions could be understood as the local
stretching induced by the map $U:{\mathcal{R}}^{3}\rightarrow{\mathcal{S}}%
^{3}$. In this real rubber-sheet geometry, the Jacobian $J$ of the
transformation provides a basic measure of the local distortion caused by the
map $U$. This enables us to build a symmetric positive definite strain tensor
defined at every point of ${\mathcal{R}}^{3}$ as
\begin{equation}
D_{ij}=J_{i}J_{j}^{T}=-\frac{1}{2}Tr\left(  L_{i}L_{j}\right)  .
\end{equation}
This strain tensor $D$, changing into $O^{T}DO$ under orthogonal
transformations, comes with three invariants expressed in terms of its
eigenvalues $\lambda_{1}^{2}$, $\lambda_{2}^{2}$ and $\lambda_{3}^{2}$:
\begin{equation}
Tr\;D=\lambda_{1}^{2}+\lambda_{2}^{2}+\lambda_{3}^{2}%
\end{equation}%
\begin{equation}
\frac{1}{2}(Tr\;D)^{2}-\frac{1}{2}Tr\;D^{2}=\lambda_{1}^{2}\lambda_{2}%
^{2}+\lambda_{2}^{2}\lambda_{3}^{2}+\lambda_{1}^{2}\lambda_{3}^{2}%
\end{equation}%
\begin{equation}
det\;D=\lambda_{1}^{2}\lambda_{2}^{2}\lambda_{3}^{2}.
\end{equation}
Since it is assumed that geometrical distorsion is unaffected by rotations of
the coordinates frame in both space and isospace, the energy density should
remain invariant and could be written as a function of the basic invariants as
follows
\begin{equation}
{\mathcal{E}}=\alpha\left(  \lambda_{1}^{2}+\lambda_{2}^{2}+\lambda_{3}%
^{2}\right)  +\beta\left(  \lambda_{1}^{2}\lambda_{2}^{2}+\lambda_{2}%
^{2}\lambda_{3}^{2}+\lambda_{1}^{2}\lambda_{3}^{2}\right)  \label{Energy}%
\end{equation}
where $\alpha$ and $\beta$ are parameter depending on $F_{\pi}$ and $e,$ while
the baryon density is associated with the quantity
\begin{equation}
B^{0}=\frac{\lambda_{1}\lambda_{2}\lambda_{3}}{2\pi^{2}}.
\end{equation}
In this picture, radial strains are orthogonal to angular ones. Moreover,
owing to the conformal aspect of $R_{N}(z)$, angular strains are isotropic.
Thereby, it is customary to identify
\begin{equation}
\lambda_{1}=-F^{\prime}(r)
\end{equation}
and
\begin{equation}
\lambda_{2}=\lambda_{3}=\frac{\sin F(r)}{r}\ \frac{1+|z|^{2}}{1+|R_{N}%
(z)|^{2}}\ \left\vert \frac{dR_{N}(z)}{dz}\right\vert \ .
\end{equation}
Thus, substituting these eigenvalues in (\ref{Energy}) and integrating over
physical space yields
\begin{equation}
M_{N}=\int dr\left(  \frac{\pi F_{\pi}^{2}}{2}\;{\mathcal{E}}_{2}^{N}%
+\frac{2\pi}{e^{2}}\;{\mathcal{E}}_{4}^{N}\right)
\end{equation}
with
\begin{equation}
{\mathcal{E}}_{2}^{N}=\left(  F^{\prime2}+2N\frac{\sin^{2}F}{r^{2}}\right)
\end{equation}%
\begin{equation}
{\mathcal{E}}_{4}^{N}=\left(  2NF^{\prime2}\frac{\sin^{2}F}{r^{2}%
}+{\mathcal{I}}_{N}\frac{\sin^{4}F}{r^{2}}\right)
\end{equation}
and
\begin{equation}
{\mathcal{I}}_{N}=\frac{1}{4\pi}\int\frac{2idzd\bar{z}}{|1+|z|^{2}|^{2}%
}\ \left(  \frac{1+|z|^{2}}{1+|R_{N}(z)|^{2}}\left\vert \frac{dR_{N}(z)}%
{dz}\right\vert \right)  ^{4},
\end{equation}
wherein $\frac{2idzd\bar{z}}{|1+|z|^{2}|^{2}}$ corresponds to the usual area
element on a 2-sphere, that is $\sin\theta\ d\theta\ d\phi$. At first glance,
one sees that radial and angular contributions to the static energy are
clearly singled out.

Now, focussing on the $N=2$ case, the most general rational map reads
\begin{equation}
R_{2}(z)=\frac{\alpha z^{2}+\beta z+\gamma}{\mu z^{2}+\nu z+\lambda}.
\end{equation}
However, imposing the exact $2-$torus symmetries (axial symmetry and rotations
of $180^{o}$ around Cartesian axes), as expected from numerical analysis
\cite{Verbaarschot}, restricts the general form above to this one
\begin{equation}
R_{2}(z)=\frac{z^{2}-a}{-az^{2}+1}.
\end{equation}
It has been showed in \cite{Houghton} that ${\mathcal{I}}_{2}$, and thus the
mass functional alike, exhibits a minimum for $a=0$. Then, one must conclude
that the most adequate choice of a rational map for the description of the
biskyrmion solution boils down to $R_{2}(z)=z^{2}$. Recasting this map in term
of angular variables $(\theta,\phi)\rightarrow(\Theta,\Phi)$, we get
\begin{align*}
z  &  \rightarrow R_{2}(z)=\tan(\frac{\Theta}{2})\ e^{i\Phi}=\tan^{2}%
(\frac{\theta}{2})\ e^{2i\phi}\\
\theta &  \rightarrow\Theta(\theta)=\arcsin\left(  \frac{\sin^{2}\theta
}{1+\cos^{2}\theta}\right) \\
\phi &  \rightarrow\Phi(\phi)=2\phi
\end{align*}
It is easy to verify that for $B=1,$ the rational map is simply $R(z)=z,$ or
$\Theta(\theta)=\theta$ and $\Phi(\phi)=\phi$, and one recovers the energy
density in \cite{Houghton}.

We shall assume here that this angular function still holds in the oblate
picture in (\ref{m2}-\ref{K2}). After analytical angular integrations are
performed, the mass of the oblate biskyrmion can be cast in the form
\begin{equation}
M=\frac{4\pi\epsilon}{\lambda}\int_{0}^{\infty}d\eta\left(  \frac{\tilde{d}%
}{2}\;{\mathcal{E}}_{2}+\frac{1}{4\tilde{d}}\;{\mathcal{E}}_{4}\right)
\end{equation}
with
\begin{equation}
{\mathcal{E}}_{2}={\mathcal{M}}_{21}f^{\prime2}+{\mathcal{M}}_{22}\sin^{2}f
\end{equation}
and
\begin{equation}
{\mathcal{E}}_{4}={\mathcal{M}}_{41}f^{\prime2}\sin^{2}f+{\mathcal{M}}%
_{42}\sin^{4}f.
\end{equation}

The explicit expressions of the density functions ${\mathcal{M}}_{ij}(\eta)$,
which are reported in Appendix A, follow from straightforward but tedious
calculations. Adopting the same conventions as in \cite{Marleau}, for the sake
of comparison, we also rescaled the deformation parameter as ${\tilde{d}%
}=eF_{\pi}d/2\sqrt{2}$, with $\epsilon=1/\sqrt{2}e$ and $\lambda=2/{F_{\pi}}$.
The values of $F_{\pi}=129$ MeV and $e=5.45$ are chosen to coincide with those
of \cite{Adkins}. Now, the chiral angle function $f(\eta)$ can be determined
from the minimization of the above functional, i.e. requiring that $\delta
M[f(\eta)]=0$ . Thus static field configuration must obey the following
nonlinear second-order ODE:
\begin{align}
0  &  =f^{\prime\prime}\left(  2\tilde{d}\cosh\eta+\frac{1}{2\tilde{d}%
}{\mathcal{M}}_{41}\sin^{2}f\right)  +f^{\prime2}\left(  \frac{1}{2\tilde{d}%
}{\mathcal{M}}_{41}\sin f\cos f\right) \nonumber\\
&  +f^{\prime}\left(  2\tilde{d}\sinh\eta+\frac{1}{2\tilde{d}}{\mathcal{M}%
}_{41}^{\prime}\sin^{2}f\right) \nonumber\\
&  -\tilde{d}{\mathcal{M}}_{22}\sin f\cos f-\frac{1}{\tilde{d}}{\mathcal{M}%
}_{42}\sin^{3}f\cos f\quad.
\end{align}
Here, the primes merely denote derivatives with respect to $\eta$. Solving
numerically for several values of $\tilde{d}$, we obtain the set of chiral
angle functions of Figure 1. When $\tilde{d}$ is small, we recover exactly the
solution of the $N=2$ rational map ansatz. Let us stress that increasing
$\tilde{d}$ enforces a continuous displacement of the function $f(\eta)$ which
induces a smooth deformation of the soliton.

\begin{figure}[tbh]
\begin{center}
\includegraphics[height=6cm, width=7cm]{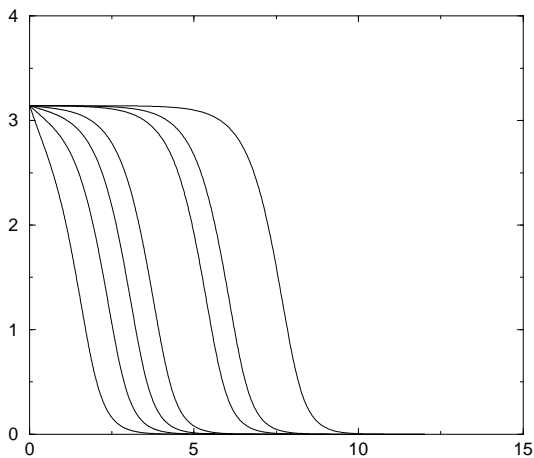}
\end{center}
\caption{Oblate biskyrmion chiral angle for several values of $\tilde{d}$.
From left to right, we have $\tilde{d}=$ 0.5, 0.2,0.1, 0.05, 0.01, 0.005 and
0.001.}%
\label{fig1}%
\end{figure}

\begin{figure}[tbh]
\begin{center}
\includegraphics[height=6cm, width=7cm]{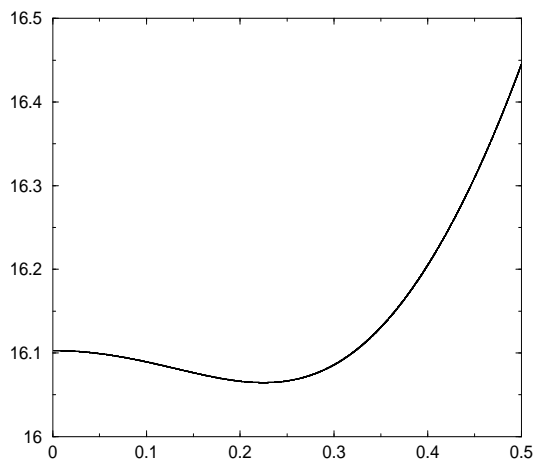}
\end{center}
\caption{ The mass of the oblate biskyrmion solution as a function of
$\tilde{d}$. The mass reaches its minimal value for $\tilde{d}=0.225$.}%
\label{fig2}%
\end{figure}In Figure 2, we plot the mass of the oblate biskyrmion as a
function of $\tilde{d}$. The mass of the biskyrmion passes trough a minimum
for a finite non-zero value of the parameter $\tilde{d}$. This is a clear
indication that the oblate solution is energically favored. Note again that in
the limit $\tilde{d}\rightarrow0$, we reproduce the mass value found in
\cite{Houghton, Krusch} with the rational maps ansatz and profile function
with radial dependence $F(r)$.

\section{NUMERICAL RESULTS AND DISCUSSION}

Numerical calculations carried out some years ago by Verbaarschot
\cite{Verbaarschot} and almost concurrently by Kopeliovich and Stern
\cite{Kopeliovich}, establish that in the Skyrme model the mass ratio
$R_{2/1}=M_{B=2}/M_{B=1}$ is $1.92$. The $B=2$ oblate solution also represents
a bound state of two solitons since its mass is lower than twice the mass
found in the $B=1$ sector \cite{Marleau}. From our calculations, we get that
the mass of the static oblate biskyrmion, being minimized for $\tilde
{d}=0.225$, is $16.066\frac{4\pi\epsilon}{\lambda}$ or $1689.5$ MeV. The
parameter $\tilde{d}=0.225$ provides a measure of how the the $B=2$ solution
is flattened. Hence, the mass ratio of our flattened solution turns out to be
$R_{2/1}=1.958$. Comparing to other ansatz for $B=2$ \ solutions, it is fairly
smaller than that predicted by the familiar hedgehog ansatz with boundary
conditions $F(0)=2\pi$ and $F(\infty)=0$, since then $R_{2/1}>3$
\cite{Jackson} or and still better than the hedgehog-like solution with
$\phi\rightarrow2\phi$ proposed as in \cite{Weigel}, whose mass ratio is
$R_{2/1}=2.14$ . Note that none of these solutions are stable solutions since
$R_{2/1}>2$. Let us mention that Kurihara et al. \cite{Kurihara} achieved a
mass ratio of $1.94$ using a different angular parametrization. However, there
are no obvious physical grounds for their angular trial function and it
remains that rational maps are far more superior when it comes to depict the
symmetries of the $B>1$ solutions. Our results still represent a slight
improvement over those obtained in the original framework of rational maps,
i.e. $R_{2/1}=1.962$ \cite{Houghton}, where the chiral angle is strictly
radial. Hence the oblate solution which depends of a spheroidal oblate
coordinate $\eta,$ captures more exactly the profile shape of the biskyrmion
than the original rational maps ansatz does although both rely on rational
maps. The relatively small improvement also suggests that a better ansatz for
classical $B=2$ static solution would require a different choice for the
angular dependence. In that regard, \cite{Houghton2} and \cite{Ioannidou} both
achieved a mass ratio of $1.933$ by dropping the constraint on the rational
maps to be holomorphic. These alternatives remain very difficult to implement
for an oblate field configuration.

It is worth emphasizing that the procedure presented here generalizes to any
baryon number and any choice of angular ansatz consistent with multiskyrmions
symmetries although analytical angular integration may become more cumbersome
if not impossible in those cases. Similarly, the approach can be generalized
to other Skyrme-like effective Lagrangians.

In this paper we only investigated the classical $B=2$ static solution but in
principle, the full solution requires a quantization treatment to account for
the quantum properties of dibaryons. The most standard procedure consists in a
semiclassical quantization using collective variables. It only adds simple
kinetic terms to the Hamiltonian but these energy contributions should
partially fill the energy gap between the $(I=0,J=1)$ deuteron mass ($1876$
MeV) and that of our deformed $B=2$ static Skyrmion ($1689.5$ MeV). So, even
if our analytical oblate ansatz is not necessarily the lowest static energy
solution for the $B=2$ Skyrmion, the optimization of the oblateness parameter
$\tilde{d}$ should prove adequate to take into account the soliton deformation
due to centrifugal effects, as for the $B=1$ case \cite{Marleau}. Thus,
following such a procedure, we can expect that the properly quantized
biskyrmion solution would provide a good starting point for the description of
the low energy phenomenology of the deuteron \cite{Braaten,Leese,Riska}. The
problem of quantization of the oblate biskyrmion solution is an important
topic in itself and will be addressed elsewhere.

\section{ACKNOWLEDGEMENTS}

J.-F. Rivard is indebted to H. Jirari for useful discussions on computational
methods and numerical analysis. This work was supported in part by the Natural
Sciences and Engineering Research Council of Canada.

\section{APPENDIX}

Performing angular integrations in the oblate static energy functional, we get
the following ${\mathcal{M}}_{ij}(\eta)$ density functions:%

\[
{\mathcal{M}}_{21}=2\cosh\eta
\]%
\[
{\mathcal{M}}_{22}=\frac{2\cosh^{2}\eta-7}{\cosh\eta}+\frac{\pi(8\cosh^{2}%
\eta-7)}{2\cosh^{2}\eta\sqrt{\cosh^{2}\eta-1}}+\frac{6\cosh^{4}\eta
-11\cosh^{2}\eta+7}{2\cosh^{2}\eta}L(\eta)
\]%
\begin{align*}
{\mathcal{M}}_{41}  &  =\frac{L(\eta)\ \left(  2\cosh^{12}\eta-15\cosh
^{10}\eta+45\cosh^{8}\eta-86\cosh^{6}\eta+124\cosh^{4}\eta-92\cosh^{2}%
\eta+24\right)  }{2(\cosh^{6}\eta-4\cosh^{4}\eta+8\cosh^{2}\eta-4)^{2}}\\
&  -\frac{\pi\ \left(  8\cosh^{10}\eta-53\cosh^{8}\eta+120\cosh^{6}%
\eta-136\cosh^{4}\eta+76\cosh^{2}\eta-16\right)  }{2\sqrt{\cosh^{2}\eta
-1}\ (\cosh^{6}\eta-4\cosh^{4}\eta+8\cosh^{2}\eta-4)^{2}}\\
&  +\frac{8\cosh^{5}\eta\ \arctan\left(  \frac{1}{\sqrt{\cosh^{2}\eta-1}%
}\right)  \ (\cosh^{6}\eta-5\cosh^{4}\eta+7\cosh^{2}\eta-3)}{\sqrt{\cosh
^{2}\eta-1}\ (\cosh^{6}\eta-4\cosh^{4}\eta+8\cosh^{2}\eta-4)^{2}}\\
&  +\frac{(2\cosh^{11}\eta-11\cosh^{9}\eta+30\cosh^{7}\eta-40\cosh^{5}%
\eta+28\cosh^{3}\eta-8\cosh\eta)}{(\cosh^{6}\eta-4\cosh^{4}\eta+8\cosh^{2}%
\eta-4)^{2}}+2L(\eta)
\end{align*}%
\begin{align*}
{\mathcal{M}}_{42}  &  =\frac{1}{4\cosh^{2}\eta}\left(  (2\cosh^{4}\eta
+\cosh^{2}\eta-7)\ L(\eta)-\frac{2\cosh\eta(2\cosh^{4}\eta-11\cosh^{2}\eta
+7)}{(\cosh^{2}\eta-1)}\right) \\
&  +\frac{\pi\ (4\cosh^{2}\eta-3)}{2\cosh^{2}\eta\sqrt{\cosh^{2}\eta-1}}.
\end{align*}
where $L(\eta) = \ln\left[  \frac{\cosh\eta+ 1}{\cosh\eta- 1} \right]  $.


\begin{thebibliography}{99}                                                                                               %


\bibitem {Skyrme}T.H.R Skyrme, Proc. R. Soc. A260, 127 (1961).\newline T.H.R
Skyrme, Proc. R. Soc. A262, 237 (1961).\newline T.H.R Skyrme, Nucl. Phys. 31,
556 (1962).

\bibitem {t'Hooft}G. t'Hooft, Nucl. Phys. B72, 461 (1974).

\bibitem {Witten}E. Witten, Nucl. Phys. B160, 57 (1979).

\bibitem {Braaten}E. Braaten and L. Carson, Phys. Rev. D 38, 3525 (1988).

\bibitem {Houghton}C.J. Houghton, N.S. Manton, and P.M. Sutcliffe, Nucl. phys.
B 510, 507 (1998).

\bibitem {Scoccola}N.N. Scoccola, arXiv:hep-ph/9911402 v1 18 Nov. 1999.

\bibitem {Adkins}G.S. Adkins, C.R. Nappi, and E. Witten, Nucl. Phys. B228, 552
(1983).\newline G.S. Adkins, C.R. Nappi, Nucl. Phys. B233, 109 (1984).

\bibitem {Jackson}A.D. Jackson and M. Rho, Phys. Rev. Lett. 51, 751 (1983).

\bibitem {Weigel}H. Weigel, B. Schwesinger, and G. Holzwarth, Phys. Lett.
168B, 321 (1986).

\bibitem {Verbaarschot}J.J.M. Verbaarschot, Phys. Lett. B 195, 235 (1987).

\bibitem {Atiyah}M.F. Atiyah and N.S. Manton, Phys. Lett. B222, 438 (1989).

\bibitem {Thomas}G.L. Thomas, N.N. Scoccola, and A. Wirzba, Nucl. Phys. A575,
623 (1994).

\bibitem {Kurihara}T. Kurihara, H. Kanada, T. Otofuji, and Sakae Saito, Prog.
Theor. Phys. 81, 858 (1989).

\bibitem {Hajduk}C. Hajduk, and B. Schwesinger, Nucl. Phys. A 453, 620 (1986).

\bibitem {Marleau}F. Leblond, and L. Marleau, Phys. Rev. D 58, 054002 (1998).

\bibitem {Houghton2}C.J. Houghton and S. Krusch, J. Math. Phys. 42, 4079 (2001).

\bibitem {Ioannidou}T. Ioannidou, B. Kleihaus, W. Zakrzewski, Phys. Lett.
B597, 546 (2004).

\bibitem {Krusch}S. Krusch, Nonlinearity 13, 2163 (2000).

\bibitem {Manton}N.S. Manton, Comm. Math. Phys. 111, 469 (1987).

\bibitem {Kopeliovich}V.B. Kopeliovich, and B.E. Stern , JETP Lett. 45, 203 (1987).

\bibitem {Leese}R.A. Leese, N.S. Manton, B.J. Schroers, Nucl. Phys. B442, 228 (1995).

\bibitem {Riska}T. Krupovnickas, E. Norvaisas and D.O. Riska, Lithuanian
Journal of Physics 41, 13 (2001), arXiv:nucl-th/0011063 v1 17 Nov 2000; A.
Acus, J. Matuzas, E. Norvaisas and D.O. Riska, arXiv:nucl-th/nucl-th/0307010
v1 2 Jul 2003.
\end{thebibliography}
\end{document}